\documentclass[aps,pra,superscriptaddress,amsmath,amssymb,preprintnumbers,showpacs,twocolumn]{revtex4}
\usepackage{amssymb}
\usepackage{graphicx}
\usepackage{bm}
\usepackage{color}
\usepackage{subfigure}

\begin{document}

\title{\small Heat Capacity and Entanglement Measure in a simple two-qubit model}

\author{B. Leggio}
\address{Dipartimento di Fisica, Universit\`{a} di Palermo, Via Archirafi 36, 90123 Palermo,
Italy}

\author{A. Napoli}
\address{Dipartimento di Fisica, Universit\`{a} di Palermo, Via Archirafi 36, 90123 Palermo,
Italy}

\author{H. Nakazato} 
\address{Department of Physics, Waseda University, Tokyo 169-8555, Japan}

\author{A. Messina}
\address{Dipartimento di Fisica, Universit\`{a} di Palermo, Via Archirafi 36, 90123 Palermo,
Italy}

\newcommand{\ket}[1]{\displaystyle{|#1\rangle}}
\newcommand{\bra}[1]{\displaystyle{\langle #1|}}

\begin{abstract}
A simple two-qubit model showing Quantum Phase Transitions as a consequence of ground state level crossings is studied in detail. Using the Concurrence of the system as an entanglement measure and heat capacity as a marker of thermodynamical properties, an analytical expression giving the latter in terms of the former is obtained. A protocol allowing an experimental measure of entanglement is then presented and compared with a related proposal recently reported by $\mathrm{Wie\acute{s}niak}$, Vedral and Brukner
\end{abstract}
\maketitle

\section*{Introduction}
Entanglement as a strikingly non-classical feature has recently been linked to Quantum Phase Transitions (QPTs) in multipartite systems \cite{oster, osb, gre}. By definition, a quantum phase transition \cite{sac} is a sudden change of some physical properties a system exhibits at zero temperature, due to an abrupt modification of its ground state when a non-thermal parameter characterizing its Hamiltonian crosses a critical value \cite{qptgap} (known as quantum critical point or QCP). There exist two main classes of QPTs: the first type, known as First Order QPT (I-QPT), is a consequence of the occurrence of a level crossing \cite{vedrlevcro} in the ground state energy in correspondence of the quantum critical point. This leads to a jump in the derivative of the energy of the system and to a discontinuity in the order parameter characterizing the transition. The second kind of QPTs, known as Second Order QPT (II-QPT) \cite{vidalpal}, can occur only in the thermodynamic limit when the dimension of the involved Hilbert space goes to infinity. In this case the quantum critical point is the limit of an avoided level crossing as the number of particles diverges. In both cases however the QCP manifests itself as a non-analytical point in the ground state energy. Consequences of such a non-analyticity, in some particular cases, might result in some peculiar properties of the system even at finite temperatures.\\
A great effort has thus been devoted to the study of a particular kind of entanglement, namely the so-called thermal entanglement \cite{therment}. Its behavior, when non-thermal parameters are varied, is a signature of the physics underlying the transition itself \cite{lutian}. It has been shown \cite{wu} that, under appropriate conditions, the occurrence of a quantum critical point against a non-thermal parameter $\kappa$, at $\kappa = \bar{\kappa}$, is a necessary and sufficient condition for the existence of a discontinuity at $\bar{\kappa}$ in some entanglement measures or in their derivatives. \\
Moreover the renewed interest towards quantum thermodynamics in the last years \cite{quanther, hart} has spurred to search a possible link between the entanglement exhibited by a physical system in a thermal state and its thermal properties \cite{liao, banuls, vedr}. \\
In view of these considerations it seems interesting to investigate the possibility of using a thermal quantity, like for example heat capacity, as a marker of a wide class of quantum phase transitions and to give a direct link between such a quantity and thermal entanglement in the state of the system under scrutiny. \\
Some results in this direction have already been obtained \cite{heatent} but a full clear picture is not yet available. The aim of this work is to show that such an explicit link can be found, at least in a simple system of two spin-$\frac{1}{2}$ particles.\\
This paper is organized as follows. In Section 1 we study a simple model of two qubits interacting with Heisenberg-like exchange potential and we define thermal quantities and entanglement properties for a thermal state of such a system. The main result of this paper is shown in Section 2, namely a direct link between a measurable thermal quantity and an entanglement measure, showing how this link can be employed to define an entanglement measuring protocol. Finally a comparison between our results and a similar (but not equivalent) result recently obtained by $\mathrm{Wie\acute{s}niak}$, Vedral and Brukner \cite{vedr} is performed.\\

\section{Hamiltonian Model}
One of the most easily solvable spin models is the one described by a Heisenberg-like spin-spin interaction \cite{heis}. Its Hamiltonian characterizes many real physical situations and it is perhaps the simplest Hamiltonian operator showing quantum critical points \cite{critdyn, ren}.\\
Nevertheless, when the number of atoms grows an exact diagonalization of the corresponding Hamiltonian becomes more and more difficult, if possible at all. Even in a few cases where exact results are available in the thermodynamic limit \cite{pater}, the dimension of involved Hilbert spaces is so high that analytical results are usually hard to be exploited in order to perform a detailed study \cite{chall}. Since the knowledge of the Hamiltonian spectrum and the partition function will be of fundamental importance throughout our work and since our goal is not a study of full phase diagrams for macroscopic systems, we will deal with a simple system composed of two qubits only. Moreover we will work with a simplified XX interaction term \cite{wang} in order to obtain a particularly easy form for Hamiltonian spectrum. \\
The Hamiltonian then reads
\begin{equation}\label{ham1}
H=-\frac{\lambda}{2}(\sigma_{1}^{x}\sigma_{2}^{x}+\sigma_{1}^{y}\sigma_{2}^{y})-\frac{h}{2}\sigma^{z}
\end{equation}
where $\lambda$ is a coupling constant describing exchange interactions between the two spins, $h$ is an external magnetic field in the $z$ direction, $\vec{\sigma}_{1}$ ($\vec{\sigma}_{2}$) is the Pauli matrices vector for the first (second) atom and $\vec{\sigma}=\vec{\sigma}_{1}+\vec{\sigma}_{2}$.\\
It is easy to show that this Hamiltonian can be cast in the form
\begin{equation}\label{ham2}
H=-\lambda(S^{2}-S_{z}^{2}-I)-hS_{z}
\end{equation}
where $\vec{S}=\frac{\vec{\sigma}}{2}$. The advantage of working with $H$ as given by (\ref{ham2}) is that its diagonalization is now straightforward, since this Hamiltonian is diagonal in the coupled basis $\Big\{|S,M\rangle\Big\}=\Big\{|1,1\rangle,|1,0\rangle,|1,-1\rangle,|0,0\rangle\Big\}$ of common eigenvectors of $S^{2}$ and $S_{z}$.\\
In this ordered basis $H$ reads
\begin{eqnarray} \label{spec}
H=\left(\begin{array}{cccc}
-h & 0 & 0 & 0 \\
0 & -\lambda & 0 & 0 \\
0 & 0 & h & 0 \\
0 & 0 & 0 & \lambda \\
\end{array}\right).
\end{eqnarray}\\
\subsection{Thermal and entanglement properties, and thermal state of the system}
Starting from the knowledge of Hamiltonian eigenvalues it is possible to obtain a closed form for the partition function $Z$ of the system which reads
\begin{equation}\label{part}
Z=2\cosh(\beta h)+2\cosh(\beta \lambda).
\end{equation}
Here $\beta=\frac{1}{k_{B}T}$.\\
In view of our target to link thermodynamical quantities to entanglement measures, equation (\ref{part}) is of great importance since it encompasses all thermodynamical properties of our system. We are now able to give an expression of the heat capacity. Since ($k_{B}=1$)
\begin{equation}
U=-\frac{\partial \ln(Z)}{\partial \beta}=-\frac{h \sinh(\beta h)+\lambda \sinh(\beta \lambda)}{\cosh(\beta h)+\cosh(\beta \lambda)}
\end{equation}
we obtain
\begin{eqnarray}\label{heatcap}
&\frac{C_{V}}{\beta^{2}}=-\frac{\partial U}{\partial \beta}=\nonumber \\
&\frac{h^{2}+\lambda^{2}+\frac{1}{2}(\lambda - h)^{2}\cosh\big(\beta(\lambda + h)\big)+\frac{1}{2}(\lambda + h)^{2}\cosh\big(\beta(\lambda - h)\big)}{(\cosh(\beta h)+\cosh(\beta \lambda))^{2}}.
\end{eqnarray} \\
What we are interested in is the study of a thermal state of our system whose density matrix $\rho$ can be cast as follows:
\begin{equation} \label{thermalstate}
\rho=\frac{1}{Z}e^{-\beta H},
\end{equation}
$Z$ and $H$ being respectively given by (\ref{part}) and (\ref{ham2}). In the coupled basis, such an operator reads
\begin{eqnarray}
\rho=\frac{1}{Z}\left(\begin{array}{cccc}
e^{\beta h} & 0 & 0 & 0 \\
0 & e^{\beta \lambda} & 0 & 0 \\
0 & 0 & e^{-\beta h} & 0 \\
0 & 0 & 0 & e^{-\beta \lambda} \\
\end{array}\right).
\end{eqnarray}
\\ \\
On the other hand we wish to quantify entanglement between two qubits in our system. In order to measure the degree of quantum correlation characterizing the two-spin system in the state given by equation (\ref{thermalstate}), it is possible to use many different parameters: one of them is the negativity of the state of the system; another typical choice is the concurrence. We will use this latter parameter thanks to its particularly simple expression in terms of Hamiltonian parameters.\\
Our entanglement measure will thus be defined as $\mathcal{C}=\mathrm{Max}\{0,\nu\}$, where $\nu=\sqrt{\mu_{1}}-\sqrt{\mu_{2}}-\sqrt{\mu_{3}}-\sqrt{\mu_{4}}$. Here the $\mu_{i}$s are eigenvalues of the matrix $\rho(\sigma_{y}\otimes\sigma_{y})\rho^{*}(\sigma_{y}\otimes\sigma_{y})$ and $\mu_{1}\geq\mu_{2}\geq\mu_{3}\geq\mu_{4}$. Evaluating such a quantity, after some trivial calculations, we obtain
\begin{equation}\label{conc}
\mathcal{C}=\mathrm{Max}\Big\{ 0,\nu=\frac{2}{Z}(|\sinh{\beta \lambda}|-1) \Big\}.
\end{equation}
We notice that, strictly speaking, the system in its thermal state is entangled if and only if $|\lambda| > \frac{1}{\beta}\ln{(1+\sqrt{2})}=\bar{\lambda}$. One could be surprised noticing that such a critical value for $\lambda$ does not depend upon $h$. Nevertheless the partition function explicitly depends upon $h$ in such a way that, when $h \gg \lambda$, $\mathcal{C}$ is almost zero even if $\lambda > \bar{\lambda}$. Only when $\lambda \gtrsim h$ the concurrence is significantly different from zero. Physically this statement means that entanglement can arise between the two spins only when their mutual interaction is stronger than the magnetic one, since a strong magnetic field (with respect to the exchange interaction) forces both spins to align with it, erasing all correlations between them.\\
From equation (\ref{conc}) the concurrence is given in terms of three parameters, namely $\beta, \lambda$ and $h$. One way to proceed is to look at $h$ as a parameter characterizing a class of functions $\mathcal{C}_{h}(\beta, \lambda)$. In this way it is possible to find analytical expressions for the functions $\lambda_{h}(\mathcal{C}, \beta)$, valid at least in a range of values of $\mathcal{C}$.\\
What we will indeed be able to do is to find analytical expressions for the functions $\lambda_{h}(\nu, \beta)$ each of which, as long as $\nu\geq 0$, coincides with $\lambda_{h}(\mathcal{C}, \beta)$.
\section{Heat Capacity versus Concurrence: analytical results and discussion}
Starting from equation (\ref{conc}) it is possible to obtain a class of inverse functions parameterized by $h$ giving $\lambda$ in terms of $\beta$ and $\nu$. To this end we write such a set of functions as
\begin{equation}
\mathcal{C}_{h}=\mathrm{Max}\Big\{ 0,\nu=\frac{1}{\cosh{\beta h}+\sqrt{1+\sinh^{2}{\beta \lambda}}}(|\sinh{\beta \lambda}|-1) \Big\}.
\end{equation}
As we can see from (\ref{conc}), and as expected from the form of the spectrum (equation (\ref{spec})), entanglement is an even function of $\lambda$. This is due to the fact that the two eigenvectors $|10\rangle$ and $|00\rangle$ associated to the eigenvalues $\pm \lambda$ manifest the same degree of quantum correlations.\\
If we then set $x=\sinh{\beta \lambda}$ and $a=\cosh{\beta h}$ and we limit our analysis to positive values of $\lambda$ we obtain for $\nu$:
\begin{equation}
\nu=\frac{x-1}{a+\sqrt{1+x^{2}}}.
\end{equation}
It is then straightforward to obtain
\begin{equation}\label{lambdaconc}
\lambda_{h}(\nu, \beta)=\frac{1}{\beta}\mathrm{ArcSinh}\Big( \frac{a\nu+1+\nu\sqrt{(a\nu+1)^{2}+1-\nu^{2}}}{1-\nu^{2}} \Big).
\end{equation}
As said, as long as we study such a function in the range $\nu\in [0,1]$ we actually obtain the exact dependence of $\lambda$ upon $\mathcal{C}$. Nevertheless, as shown before, there is a whole range of values of $\lambda$ which results in a negative value of $\nu$ or, in other words, which results in a zero concurrence. In the whole range $\lambda \in [0,\frac{1}{\beta}\ln{(1+\sqrt{2})}]$ the concurrence as a function of $\lambda$ is not an invertible function. To overcome such an obstacle we concentrate ourselves on $\nu$ which, when non-negative, coincides with the real concurrence. In this sense it is possible to study how heat capacity depends upon $\nu$ in the whole range of values of $\lambda$, having in mind that the only physically relevant results can be obtained by looking at the restriction of the functions $C_{V}^{(h)}(\beta, \nu)$ to the set $\nu\in [0,1]$, while for negative values of $\nu$ nothing can be said about the relation between heat capacity and entanglement, since the latter is always equal to zero even if the former has not a constant value.\\
For fixed temperature and magnetic field the domain of the function $C_{V}^{(h)}(\beta, \nu)$ is then given by
\begin{equation}\label{domain}
\mathrm{D}=[-\frac{1}{a+1},1]
\end{equation}
as can be easily seen from (\ref{conc}) by evaluating the values of $\nu$ for $\lambda=0$ and for $\lambda \rightarrow \infty$.\\
By substituting (\ref{lambdaconc}) in (\ref{heatcap}) we finally obtain the set of functions $C_{V}^{(h)}(\beta, \nu)$ as
\begin{eqnarray}\label{cvcn}
\frac{C_{V}^{(h)}(\beta, \nu)}{\beta^{2}}&=&\frac{1}{(a+\cosh(\beta \lambda_{h}(\nu,\beta)))^{2}}\Bigg(h^{2}+\lambda_{h}^{2}(\nu,\beta)+ \nonumber \\
&+&\frac{1}{2}(\lambda_{h}(\nu,\beta) - h)^{2}\cosh\big(\beta(\lambda_{h}(\nu,\beta) + h)\big)+\nonumber \\
&+&\frac{1}{2}(\lambda_{h}(\nu,\beta) + h)^{2}\cosh\big(\beta(\lambda_{h}(\nu,\beta) - h)\big)\Bigg).\nonumber \\
\end{eqnarray}
Equation (\ref{cvcn}) is the main result of this work. It gives direct analytical expressions of a thermal parameter in terms of an entanglement measure and shows how, once the entanglement in the system is known, heat capacity has a well defined value. Unfortunately, the converse is not true. \\
In order to perform an analysis of the link between $C_{V}$ and $\mathcal{C}$ we can choose one single function out of the set $C_{V}^{(h)}(\beta, \nu)$. In other words, we have to fix the value of $h$ with respect to the other parameters involved in (\ref{ham2}). One possible choice is to measure all energies in units of $h$, thus obtaining the function $C_{V}^{(1)}(\beta, \nu)$. In Fig.1 we show the behavior of such a function for 9 different values of $\beta$.\\
\begin{figure}[h]\label{cvconc}
\begin{center}
\includegraphics[width=0.40\textwidth,
angle=0,clip=a]{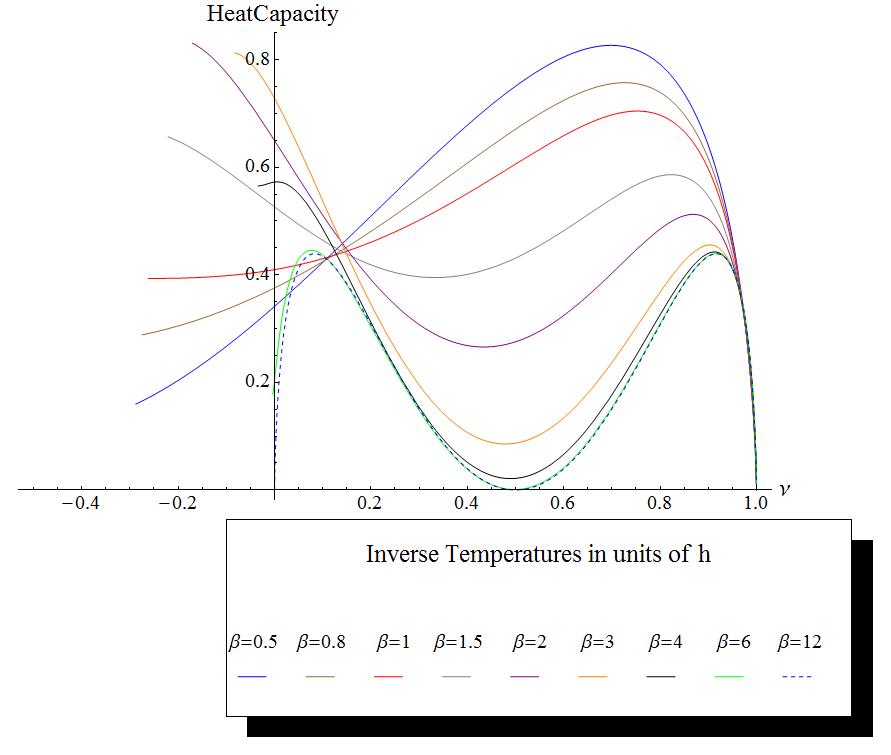}
\end{center}
\caption{Heat Capacity $C_{V}$ of the system versus Concurrence $\mathcal{C}$ (all energies are measured in units of magnetic field $h$) for 9 different values of $\beta$: $\beta=0.5$ (blue line), $\beta=0.8$ (brown line), $\beta=1$ (red line), $\beta=1.5$ (gray line), $\beta=2$ (purple line), $\beta=3$ (orange line), $\beta=4$ (black line), $\beta=6$ (green line) and $\beta=12$ (blue dashed line)}
\end{figure} \\
From this figure it is possible to notice how heat capacity always goes to zero for maximally entangled states, irrespectively of temperature. This feature may be due to the fact that a strong entanglement prevents the system from going towards excited states because of the intense correlations between its two parts. \\
Another interesting detail worth noticing is that lowering temperature results in a narrower range of negative values of $\nu$ as can be seen from equation (\ref{domain}). This means that in the limit of very low temperature the behavior of heat capacity can be fully described by changes in entanglement between the two spins in the system. This nice feature is of great interest in view of a detailed study of quantum phase transitions in spin systems. \\
Third, as it is easily seen from Fig.1, heat capacity is never a monotonic function of concurrence. Its behavior has to be analyzed in detail in order to get an insight of the physics behind it. Let us start by looking at the low temperature behavior of heat capacity. When the state of the system is separable (zero concurrence) heat capacity is zero. In Fig.2 the energy spectrum of our system is shown versus $\lambda$.
 \begin{figure}[h]\label{spectr}
\begin{center}
\includegraphics[width=0.40\textwidth,
angle=0,clip=a]{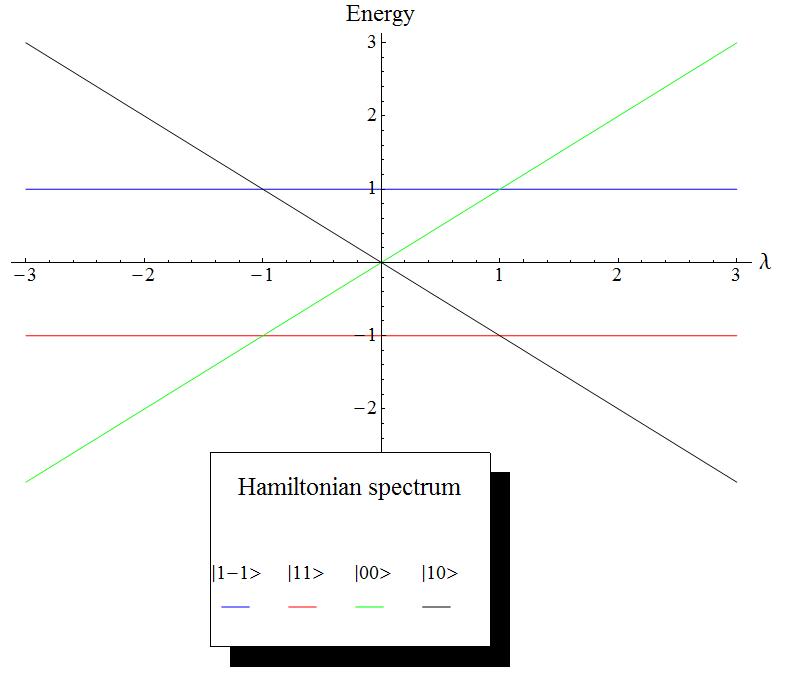}
\end{center}
\caption{Hamiltonian spectrum versus $\lambda$. All energies are measured in units of $h$}
\end{figure}
A zero value of concurrence can only be obtained for almost zero values of $\lambda$, when the ground state is a factorized one. We see from Fig.2 that for very small absolute values of $\lambda$ the gap between ground state and all the excited levels is significantly greater than the mean thermal energy when for example $\beta =12$ in Fig.1. Thus the system can not move to an excited state, resulting in a zero heat capacity. From Fig.1 we see how, moving towards greater values of concurrence, heat capacity shows two maxima and a minimum between them. Such a behavior is typical of gapped two level systems and can be seen as related to a double Schottky-like anomaly \cite{greiner, schot} right before and right after the quantum critical point. This shows how at low temperatures our system actually behaves as a two-level one. The two maxima can then be explained exploiting Schottky argument, or simply by looking at hamiltonian spectrum: raising concurrence means raising $\lambda$ ($\frac{\partial \lambda}{\partial \nu}>0$) and thus approaching the quantum critical point or, in other words, reducing the gap between ground state and first excited level. The system can now absorb thermal energy from the environment and heat capacity starts growing. Nevertheless, once $\Delta_{1} = E_{e1}-E_{g}$ becomes much smaller than $k_{B}T$ and thus the first excited level is almost as populated as the ground state one, in order to raise its energy the system has to populate the second excited level. The gap between the latter and the ground state is much greater than $\Delta_{1}$ and, at low temperatures, it is also greater than $k_{B}T$. Thus heat capacity decreases and eventually becomes zero. This explains the first maximum in Fig.1. When the critical point is crossed, the same reasoning applies and thus another maximum appears. Finally for very high values of $\lambda$ (corresponding to maximum entanglement) the gap becomes wider and wider and heat capacity rapidly goes to zero. \\
The same kind of argument can be applied to explain the high-temperature behavior of heat capacity, starting from a small but not zero value (high gap, but not much greater than mean thermal energy), growing towards a maximum (approaching the critical point) and finally decreasing to zero for high values of $\lambda$. \\
Finally, the mid-temperature heat capacity shows a peculiar behavior for small concurrence (small $\lambda$): instead of increasing with the decreasing gap between ground state and first excited levels, heat capacity decreases. This can be explained with the same argument employed to describe the minimum close to quantum critical point: when temperature is not low enough, for small $\lambda$ both the ground state and the first excited level are significantly populated. This means the system has to move to the second excited level in order to raise its energy, and the gap between this latter level and the ground state increases with increasing $\lambda$. This explains why heat capacity decreases for low concurrence.\\ \\
\section{Concurrence from Heat Capacity: an experimental protocol}
It is worth stressing once again that negative values of $\nu$ are associated with a zero concurrence and then for $\nu <0$ the behavior of heat capacity can not be described in terms of entanglement between the two spins, since this latter quantity is always zero.\\
This means that a measure of heat capacity is not able to give a definitive reply to the question about entanglement between the two particles. Let us focus for instance on the curve with $\beta =2$ (see Fig.3). It is possible to identify three different ranges in $\mathcal{C}$: the first one is associated with a negative value of $\frac{\partial C_{V}(\nu)}{\partial \nu}$, meaning that only one value of concurrence is associated to each value of heat capacity (namely, $C_{V}(\nu)$ is invertible in that range). The second one is characterized by a bouncing behavior of heat capacity and thus $C_{V}(\nu)$ is not an invertible function. Finally in the third range $C_{V}(\nu)$ is again invertible and each value of heat capacity is associated to a single value of concurrence. The existence of these ranges is a direct consequence of the behavior of the function $C_{V}(\lambda)$. Indeed, since $\frac{\partial C_{V}}{\partial \nu}=\frac{\partial C_{V}}{\partial \lambda}\frac{\partial \lambda}{\partial \nu}$, and since $\frac{\partial \lambda}{\partial \nu}>0$, the oscillating behavior of $C_{V}(\nu)$ is due to the oscillating behavior of $C_{V}(\lambda)$.\\
Let us now suppose we actually measure heat capacity for a real system described by Hamiltonian (\ref{ham1}). We suppose we can control values of temperature (by for instance keeping our system in contact with a reservoir) and of external magnetic field. Since our system is fixed, $\lambda$ has a constant value which, nevertheless, is not known to the experimentalist. We wonder to what extent it might be possible to obtain information about microscopic details of our system (concurrence, entanglement, values of $\lambda$) by measuring $C_{V}$. In order to fix ideas, let us suppose we measure all energies in units of magnetic field and we fix temperature at $\beta=2$. Finally, we measure heat capacity to get the value $C_{meas2}$ ($C_{meas2}=0.4$ in the example shown in Fig.3). It is easily seen from Fig.1 that there exists a situation in which such a value of heat capacity can be obtained with three different values of concurrence, and by this single measure there is no way we can state which one of these three values is the real one. In other words there exist three different systems, physically characterized by different values of $\lambda$ (since $h$ and $\beta$ are fixed), showing the same heat capacity. The possible values of $\lambda$ compatible with the measured value $C_{meas2}$ are easily obtainable from the knowledge of the function $C_{V}(\lambda, \beta=2)$ (\ref{heatcap}) or from the knowledge of both $C_{V}(\nu, \beta=2)$ (\ref{cvcn}) and $\lambda(\nu, \beta=2)$ (\ref{lambdaconc}).\\
\begin{figure}[h]\label{cvchanges}
\begin{center}
\includegraphics[width=0.40\textwidth,
angle=0,clip=a]{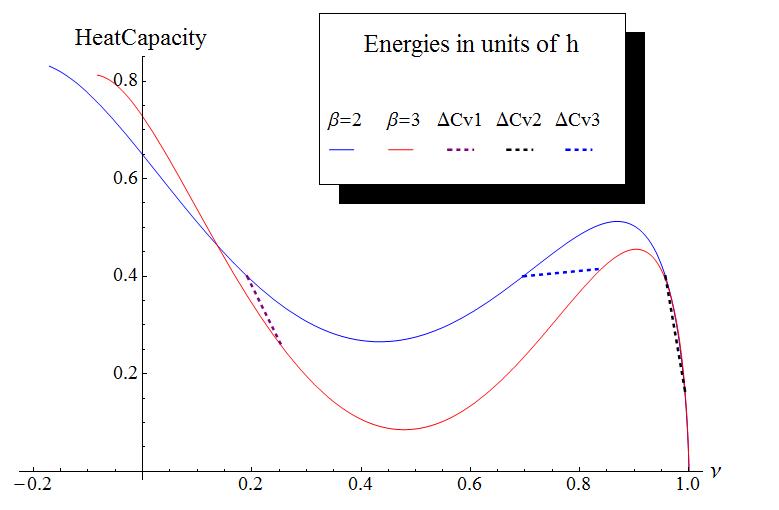}
\end{center}
\caption{Heat Capacity $C_{V}$ of the system versus parameter $\nu$ (all energies are measured in units of magnetic field $h$) for 2 different values of $\beta$: $\beta=2$ (blue line) and $\beta=3$ (red line). In figure are shown by dashed lines the changes in heat capacity and concurrence when going from $\beta =2$ to $\beta=3$ for the three possible values of $\lambda$ associated to $C_{meas2}=0.4$. It can be noticed from this figure how the changes in heat capacity strongly depend on the starting point on the curve with $\beta =2$. Heat capacity gaps $C_{V}(\beta =2)-C_{V}(\beta =3)$ are sensibly different for the three possible values of $\lambda$ discussed in the text, enabling us to distinguish between the three possible physical systems associated to them}
\end{figure}\\
We thus have three possible systems respectively characterized by values $\lambda_{1}$, $\lambda_{2}$ and $\lambda_{3}$. Looking at the example of Fig.3 with $C_{meas2}=0.4$ these three values are $\lambda_{1}=0.76054$, $\lambda_{2}=1.59078$ and $\lambda_{3}=2.68228$. The question arises whether it may be possible to identify which one of these three values is the real one.\\
Let us now suppose we measure heat capacity for the same system having a different temperature, say $\beta =3$. Since changing temperature would not affect the real value of $\lambda$ characterizing our physical system, and since this value belongs to the set $S_{a}=\big\{ \lambda_{1}, \lambda_{2}, \lambda_{3} \big\}$, we would for sure measure one of the three values $C_{V}(\lambda_{1},\beta=3)=C_{1}$, $C_{V}(\lambda_{2},\beta=3)=C_{2}$ or $C_{V}(\lambda_{3},\beta=3)=C_{3}$. Since these values are in general very different from each other it is possible for us to identify which one is the one we actually measured ($C_{meas3}$) in our second measuring process. Indeed there would be a set $S_{b}=\big\{ \lambda_{4}, \lambda_{5}, \lambda_{6} \big\}$ of values of $\lambda$ associated with the measured values $C_{meas3}$ of $C_{V}$. Following Fig.3 we might suppose the new value of heat capacity to be $C_{meas3}=0.164$ and the set $S_{b}=\big\{ 0.853576, 1.23367, 2.682 \big\}$. There will be at least one common value between $S_{a}$ and $S_{b}$. It is then possible in this way to identify the real value $\lambda_{p}$ of the coupling constant $\lambda$, since it will be the one for which both of the following equalities would hold
\begin{eqnarray}
&C_{V}&(\lambda=\lambda_{p},\beta=2)=C_{meas2} \nonumber \\
&C_{V}&(\lambda=\lambda_{p},\beta=3)=C_{meas3}. \nonumber \\
\end{eqnarray}
In the example exposed above the only common value between the sets $S_{a}$ and $S_{b}$ is found to be $\lambda_{p}=2.682$.\\
It is of course possible that more than one of the values $\big\{ \lambda_{1}, \lambda_{2}, \lambda_{3} \big\}$ is associated with the measured heat capacity $C_{meas3}$. In this case it would be enough to measure $C_{V}$ for a third temperature, to find the new set ($S_{c}$) of possible values of $\lambda$ associated with this new value of heat capacity and to look for the common element of $S_{a}$, $S_{b}$ and $S_{c}$. This procedure can be employed until the physical value of $\lambda$ is uniquely identified.\\
Once we know the actual value $\lambda_{p}$ of $\lambda$ we know everything about our system. In particular it is possible to calculate concurrence (and thus the entanglement in our system) using (\ref{conc}) which, as shown, is invertible and thus gives a unique value for $\mathcal{C}$. For the situation shown in Fig.3, exploiting the value $\lambda_{p}=2.682$, the concurrence is found to be $\mathcal{C}=0.956905$. \\
With simple measurements of heat capacity of the two-qubit system we are thus able to measure the entanglement as given by concurrence and any other microscopic property of the system. \\ \\
\section{Conclusive remarks}
It might be interesting to compare our results about heat capacity and entanglement with the one shown in \cite{vedr}, where an upper bound for heat capacity of systems in entangled states has been obtained. It is shown that there exists a maximal value of heat capacity for entangled states, such that any system showing heat capacity greater than this value is for sure in a disentangled state. Briefly, the so-called separable bound on heat capacity is a direct consequence of the existence of a separable bound on internal energy. Indeed it has been shown that for composite systems every separable state is associated to an energy greater than a certain value $E_{B}$. Moreover an explicit form for this separable bound on internal energy has been given for magnetic spin systems \cite{toth}. Applied to our system such a bound becomes, for any separable state, $U\geq E_{B}$ with
\begin{eqnarray}\label{Eb}
E_{B}=
\Bigg\{\begin{array}{cc} -|\lambda|-\frac{|h|^{2}}{4|\lambda} &  |h|\leq 2|\lambda| \\
-|h| &  |h|>2|\lambda|.
\end{array}
\end{eqnarray}
Exploiting the result (obtained in \cite{vedr}) giving the separable bound on heat capacity for gapped systems (valid for low temperature only) as
\begin{equation}\label{cvineq}
C_{V}^{Sep}\geq \beta^{2}\Delta (E_{B}-E_{g})=C_{V}^{B}
\end{equation}
where $E_{g}$ is the ground state energy, $\Delta$ is the energy gap between ground state and the first excited level and $E_{B}$ is given by (\ref{Eb}), we can obtain an expression for such a bound valid for the low temperature behavior of the system analyzed in this work, which reads
\begin{equation}\label{Cb}
\frac{C_{V}^{B}}{\beta^{2}}=\left\{
  \begin{array}{ll}
    (-\lambda -h)(-|\lambda|-\frac{|h|^{2}}{4|\lambda|}-\lambda), & \hbox{$\lambda < -h$} \\
    (\lambda +h)(-|\lambda|-\frac{|h|^{2}}{4|\lambda|}+h), & \hbox{$-h<\lambda <-\frac{h}{2}$} \\
    (\lambda +h)(-|h|+h), & \hbox{$-\frac{h}{2}<\lambda <0$} \\
    (-\lambda +h)(-|h|+h), & \hbox{$0<\lambda <\frac{h}{2}$} \\
    (-\lambda +h)(-|\lambda|-\frac{|h|^{2}}{4|\lambda|}+h), & \hbox{$\frac{h}{2}<\lambda <h$} \\
    (\lambda -h)(-|\lambda|-\frac{|h|^{2}}{4|\lambda|}+\lambda), & \hbox{$\lambda > h$}
  \end{array}\right.
\end{equation}
Notice that $C_{V}^{B}$ as a function of $\lambda$ and $h$ is always negative, while by definition heat capacity is a positive quantity. This is a consequence of the fact that energy separable bound is for our system always smaller than the ground state energy. Thus this method would give no definite reply to the question about entanglement at low temperature since (\ref{cvineq}) always holds but its violation is a sufficient but not necessary condition for the existence of quantum correlations. This shows how this latter method and the one reported in this paper, both based on measures of heat capacity, are not equivalent. It is worth noticing that, while the method based on separable bounds is (when possible) only able to detect the presence of entanglement, the one shown in this work is able to measure entanglement by means of an (indirect) measure of concurrence. \\ It is worth stressing that the low temperature behavior of entanglement is the one characterizing the critical behavior of the system and is thus of great interest; moreover when the temperature is high enough, thermal entanglement is usually destroyed. Thus the comparison presented above, even if valid only for low temperature, is yet full of physical meaning.

\section*{Acknowledgements}
One of the authors (H.N.) acknowledges the support from JSPS, Japan through a Grant-in-Aid for Scientific Research (C).


\begin{thebibliography}{25}
\bibitem{schot} A Schottky anomaly (also known as Schottky heat capacity) is the behavior of heat capacity versus temperature, typical of a two level system, characterized by a central maximum followed by an asymptotic decreasing toward zero; the only difference here is that temperature is fixed and the only parameter being varied is $\lambda$.
\end{thebibliography}
\end{document}